\documentclass[letterpaper]{article} 
\usepackage{aaai25}  
\usepackage{times}  
\usepackage{helvet}  
\usepackage{courier}  
\usepackage[hyphens]{url}  
\usepackage{graphicx} 
\usepackage{natbib}  
\usepackage{caption} 
\usepackage{algorithm}
\usepackage{algorithmic}

\usepackage{newfloat}
\usepackage{listings}
\DeclareCaptionStyle{ruled}{labelfont=normalfont,labelsep=colon,strut=off} 
\floatstyle{ruled}
\newfloat{listing}{tb}{lst}{}
\floatname{listing}{Listing}
\usepackage{times}
\usepackage{soul}
\usepackage{url}
\usepackage[utf8]{inputenc}
\usepackage{booktabs}
\usepackage{amssymb}
\usepackage{amsmath}

\usepackage{colortbl}
\definecolor{lightblue}{rgb}{0.88, 0.96, 1}
\newcommand{\brow}{\rowcolor{lightblue}}

\usepackage{latexsym}
\usepackage{enumitem}
\usepackage{color}

\title{JEN-1 Composer: A Unified Framework for \\ High-Fidelity Multi-Track Music Generation}
\author{
    Yao Yao,
    Peike Li,
    Boyu Chen,
    Alex Wang
}
\affiliations{
    Jen Music AI\\

    alex@jenmusic.ai
}

\usepackage{bibentry}

\begin{document}

\maketitle

\begin{abstract}
With rapid advances in generative artificial intelligence, the text-to-music synthesis task has emerged as a promising direction for music generation. Nevertheless, achieving precise control over multi-track generation remains an open challenge. While existing models excel in directly generating multi-track mix, their limitations become evident when it comes to composing individual tracks and integrating them in a controllable manner. This departure from the typical workflows of professional composers hinders the ability to refine details in specific tracks. To address this gap, we propose JEN-1 Composer, a unified framework designed to efficiently model marginal, conditional, and joint distributions over multi-track music using a single model. Building upon an audio latent diffusion model, JEN-1 Composer extends the versatility of multi-track music generation. We introduce a progressive curriculum training strategy, which gradually escalates the difficulty of training tasks while ensuring the model's generalization ability and facilitating smooth transitions between different scenarios. During inference, users can iteratively generate and select music tracks, thus incrementally composing entire musical pieces in accordance with the Human-AI co-composition workflow. Our approach demonstrates state-of-the-art performance in controllable and high-fidelity multi-track music synthesis, marking a significant advancement in interactive AI-assisted music creation. Our demo pages are available at \url{www.jenmusic.ai/research}.
\end{abstract}

%

\section{Introduction}
The rapid evolution of generative modeling has positioned AI-driven music generation as a prominent field, merging research innovation with practical applications in the music industry. 
Early systems like Music Transformer~\cite{huang2018music} and MuseNet~\cite{payne2019musenet}, which utilized symbolic representations~\cite{engel2017neural}, were pivotal in translating textual descriptions into MIDI-style outputs.  Although these methods were groundbreaking, their dependence on predefined virtual synthesizers often compromised the audio quality and restricted the diversity of their musical outputs.

Recent advancements in text-to-music synthesis, as demonstrated by models like MusicGen~\cite{copet2023simple}, MusicLM~\cite{agostinelli2023musiclm}, and Jen-1~\cite{li2023jen}, represent a significant leap forward in directly generating authentic audio waveforms from textual prompts. These innovations have greatly expanded the versatility and diversity of music generation, bypassing the need for extensive musical theory knowledge and traditional symbolic representations. 
However, their focus on producing composite audio mixes rather than discrete, manipulable tracks limits the level of creative control necessary in professional music production environments. Similarly, the introduction of digital audio workstations and the expansion of available timbres have indeed revolutionized musical creativity---enabling composers to explore complex harmonies, melodies, and rhythms beyond the confines of physical instruments~\cite{zhu2020pop}---but these tools still impose significant barriers. Despite their advancements, they require a deep understanding of music theory and proficiency in symbolic musical notation, which continue to pose challenges for many aspiring musicians and composers.

In response to these challenges, we propose JEN-1 Composer, a framework designed to democratize music production by streamlining the creative process. This framework employs end-to-end training to intuitively grasp the relationships between different tracks, enabling audio-to-audio orchestration that learns directly from waveform datasets. By allowing both direct audio and text input, JEN-1 Composer not only simplifies user interaction but also expands creative freedom through detailed manipulation of individual tracks.

JEN-1 Composer is a comprehensive generative framework designed to model the marginal, conditional, and joint distributions of multi-track music within a single model.
By leveraging the Jen-1~\cite{li2023jen} audio latent diffusion model as a foundation, our approach effectively and efficiently manages these distributions concurrently. 
To extend the capabilities of Jen-1, we introduce several key enhancements:
(a) We design specialized input-output configurations to handle latent representations of multiple music tracks, allowing the model to effectively capture the temporal relationships and harmonic coherence across these tracks. 
(b) We incorporate timestep vectors to govern the generation of individual tracks, providing the necessary flexibility for fine-grained control during the generation process.
(c) We augment conventional text prompts with prefix prompts to clearly define generation tasks, reducing ambiguity and improving model performance.
Additionally, we employ a curriculum training strategy that gradually introduces more complex tasks, from generating single tracks to orchestrating intricate combinations of multiple tracks.

\begin{figure*}[tb!]
    \centering
    \includegraphics[width=0.75\textwidth]{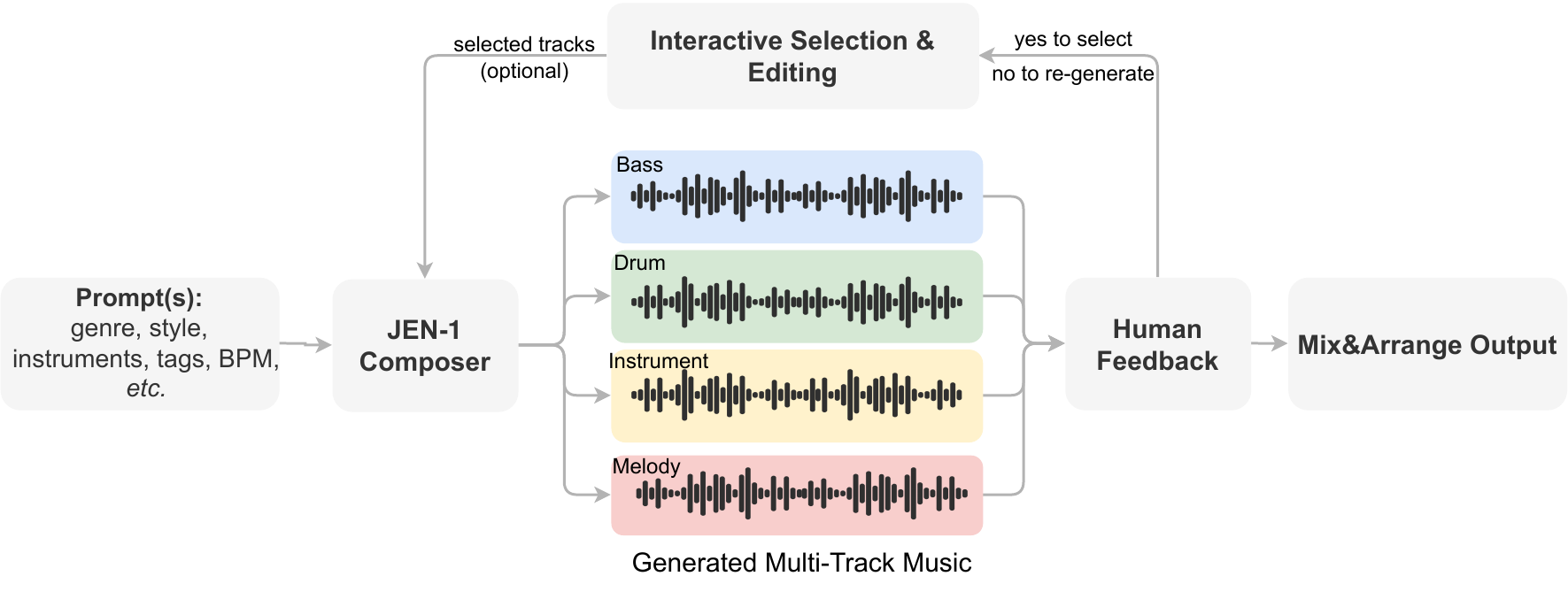}
    \caption{
The Human-AI co-composition workflow of JEN-1 Composer. JEN-1 Composer generates multiple music tracks based on user-provided text prompts (specifying genres, eras, rhythms, \textit{etc.}) and optional audio feedback, where users can select, edit, or upload tracks. The human feedback guides the generation of target tracks, ensuring temporal alignment and musical coherence. The iterative process of human feedback and AI generation continues until a harmonious and cohesive musical piece is achieved.
    }
    \label{fig:overview}
\end{figure*}

To further bridge the gap between AI capabilities and human creativity, we introduce a Human-AI co-composition workflow, as illustrated in Figure~\ref{fig:overview}. This approach enables iterative refinement of music tracks during the model's inference phase, allowing producers to collaboratively adjust and blend AI-generated tracks to align with their creative visions. Through this workflow, users can directly influence the music generation process by providing feedback on textual prompts and previously generated music tracks, ensuring that all tracks meet their precise standards.

In summary, the contributions of this work are four-fold:
\begin{enumerate}
\item We introduce a collaborative music generation workflow that seamlessly integrates human creativity with AI, designed for the iterative creation of multi-track music within an audio-based framework.
\item We present JEN-1 Composer, a unified framework that effectively models marginal, conditional, and joint distributions for multi-track music generation using a single audio latent diffusion model.
\item We design an intuitive curriculum training strategy that progressively enhances the model's capability to generate complex musical compositions.
\item We provide comprehensive quantitative and qualitative evaluations demonstrating that JEN-1 Composer achieves state-of-the-art performance in generating diverse track combinations, advancing the flexibility and creativity of music production.
\end{enumerate}

\section{Related Work}
In this section, we discuss the research background relevant to our work, focusing on two key areas: conditional music generation and multi-track music generation.

\subsection{Conditional Music Generation}
Conditional music generation has become a cornerstone in the field, with models incorporating a range of inputs, from low-level control signals like lyrics~\cite{yu2021conditional} and MIDI sequences~\cite{muhamed2021symbolic} to high-level abstract representations such as text~\cite{kreuk2022audiogen,agostinelli2023musiclm,liu2023audioldm} and images~\cite{huang2023make}. While these conditioning methods enable tailored outputs, they also present challenges in data alignment and model training. The scarcity of well-aligned data has spurred advances in self-supervised learning~\cite{marafioti2019context,borsos2023audiolm} to enhance model generalization across diverse musical contexts. Due to the complexity of raw audio waveforms, direct generation is often impractical~\cite{garbacea2019low}, leading researchers to develop feature extraction and representation strategies. Techniques like VQ-VAE and VQ-GAN, using mel-spectrograms~\cite{van2017neural,creswell2018generative,huang2023noise2music}, and quantization-based methods that transform waveforms into compact representations~\cite{zeghidour2021soundstream,defossez2022high}, have been particularly influential.

The emergence of non-autoregressive models, particularly diffusion models~\cite{ho2020denoising}, has significantly advanced the field. Models like MeLoDy~\cite{lam2023efficient} and Jen-1~\cite{li2023jen,chen2024jen} have demonstrated exceptional capabilities in producing high-fidelity music. Building on these advancements, our JEN-1 Composer integrates high-level textual prompts and inter-track dependencies to enhance alignment and harmonic cohesion among tracks, allowing for sophisticated control and compositional flexibility. Unlike previous models that typically produce composite mixes, JEN-1 Composer supports track-wise generation and iterative refinement, aligning with real-world music production workflows and fostering collaboration between humans and AI.

\subsection{Multi-track Music Generation}
Multi-track music generation enables the simultaneous creation of interdependent tracks, offering greater complexity and cohesion in music composition. Early models, such as MuseGAN~\cite{dong2018musegan}, leveraged GANs but faced issues with training instability, limited diversity, and suboptimal sound quality. Subsequent advancements, like MIDI-Sandwich2~\cite{liang2019midi} with hierarchical RNNs and VAEs, and transformer-based models such as MMM~\cite{ens2020mmm} and MTMG~\cite{jin2020transformer}, improved inter-track dependency modeling. More recently, MTT-GAN~\cite{jin2022transformer} integrated GANs with transformers to enhance adherence to musical rules.
Traditional approaches primarily relied on symbolic representations like MIDI, which constrained their ability to capture nuanced audio textures. Recent methods shift to direct audio modeling to avoid fidelity losses from intermediate representations. For example, StemGen~\cite{parker2024stemgen} uses transformers for single-stem generation, while Stable Audio~\cite{evans2024fast} and Diff-A-Riff~\cite{nistal2024diff} adopt cross-attention for prompt-based conditioning but lack multi-track alignment. Models like MSDM~\cite{marianimulti} handle multi-track spectrogram generation but do not support prompt conditioning.

Our JEN-1 Composer addresses these challenges through a diffusion-based framework designed for multi-track audio generation in the audio latent space. By directly modeling audio, it overcomes the fidelity loss inherent in symbolic and spectrum-based methods, capturing richer and more nuanced sound textures. The model leverages concatenation-based alignment to ensure inter-track coherence and incorporates text-prompt conditioning, enabling a flexible and interactive human-AI co-creation workflow. This approach surpasses traditional symbolic and spectrum-based methods in both audio fidelity and creative versatility~\cite{kong2020diffwave,liu2023audioldm}.

\section{Preliminary}
\subsection{Diffusion Model}
Diffusion models~\cite{ho2020denoising} are generative models that produce high-quality samples through iterative denoising. The process begins by gradually corrupting the original data $x_0$ with Gaussian noise over a series of timesteps in a forward process, where each noisy sample $x_t$ is generated as:
\begin{equation}
\label{latent:diff}
x_t = \sqrt{\bar{\alpha}_t}x_0+\sqrt{1-\bar{\alpha}_t}\epsilon_t,
\end{equation}
with $\epsilon_t$ as the standard Gaussian noise, $\bar{\alpha}_t=\prod^t_{i=1}\alpha_i$, and $\alpha_t=1-\beta_t$, where the sequence of $\beta_t$ is the noise schedule that controls the level of corruption over time.

In the reverse process, the diffusion model aims to recover $x_0$ by iteratively denoising $x_t$. A noise prediction model, parameterized by $\theta$, is trained to estimate the noise $\epsilon_t$ in $x_t$ at each timestep $t$ by minimizing the following loss function:
\begin{equation}
 \min_\theta \mathbb{E}_{t,x_0,\epsilon_t}\left\Vert \epsilon_t- \epsilon_\theta\left(x_t,t \right)  \right\Vert^2_2,
 \end{equation}
where $t$ is uniformly sampled from $\{1, 2, \ldots, T\}$.
With the optimized noise predictor, $x_0$ can be approximated by sampling from a Gaussian model $p\left(x_{t-1}\mid x_t\right)=\mathcal N\left(x_{t-1}\mid \mu_t\left(x_t\right), \sigma^2_t{I}\right)$ in a stepwise manner~\cite{bao2023one}. The optimal mean for this Gaussian, under maximum likelihood estimation, is:
\begin{equation}
    \mu^*_t\left(x_t\right) = \frac{1}{\sqrt{\alpha}_t}\left(x_t-\frac{\beta_t}{\sqrt{1-\bar{\alpha}_t}}\epsilon_\theta\left(x_t,t \right)\right).
\end{equation}
By iteratively applying this process, the diffusion model refines the noise, progressively generating new samples that closely resemble the original training data.

\subsection{Audio Latent Diffusion Model}
Modeling raw audio waveforms directly poses challenges due to their high dimensionality,
where $x_0 \in \mathbb{R}^{C \times S}$ represents the waveform, with $C$ denoting the number of channels and $S$ indicating the sequence length.
To address this, Jen-1~\cite{li2023jen} extends the Latent Diffusion Model (LDM)~\cite{rombach2022high} framework, originally formulated for images, to the domain of audio generation.
In the Jen-1 architecture, the audio waveform $x_0$ is mapped to a lower-dimensional latent representation $z_0 \in \mathbb{R}^{D \times S^{\prime}}$ via an audio encoder $f_\phi$, and then reconstructed back to the original waveform $\widehat{x}_0$ through an audio decoder $g_\psi$. Here $S^{\prime} \ll S$ is the compressed sequence length and $D$ is the latent dimension. This process is denoted as:
\begin{equation}
\label{eq:latent}
z_0 = f_{\phi}(x_0), \quad \widehat{x}_0 = g_{\psi}(z_0) \approx x_0.
\end{equation}

\begin{figure}[tb!]
    \centering
    \includegraphics[width=0.48\textwidth]{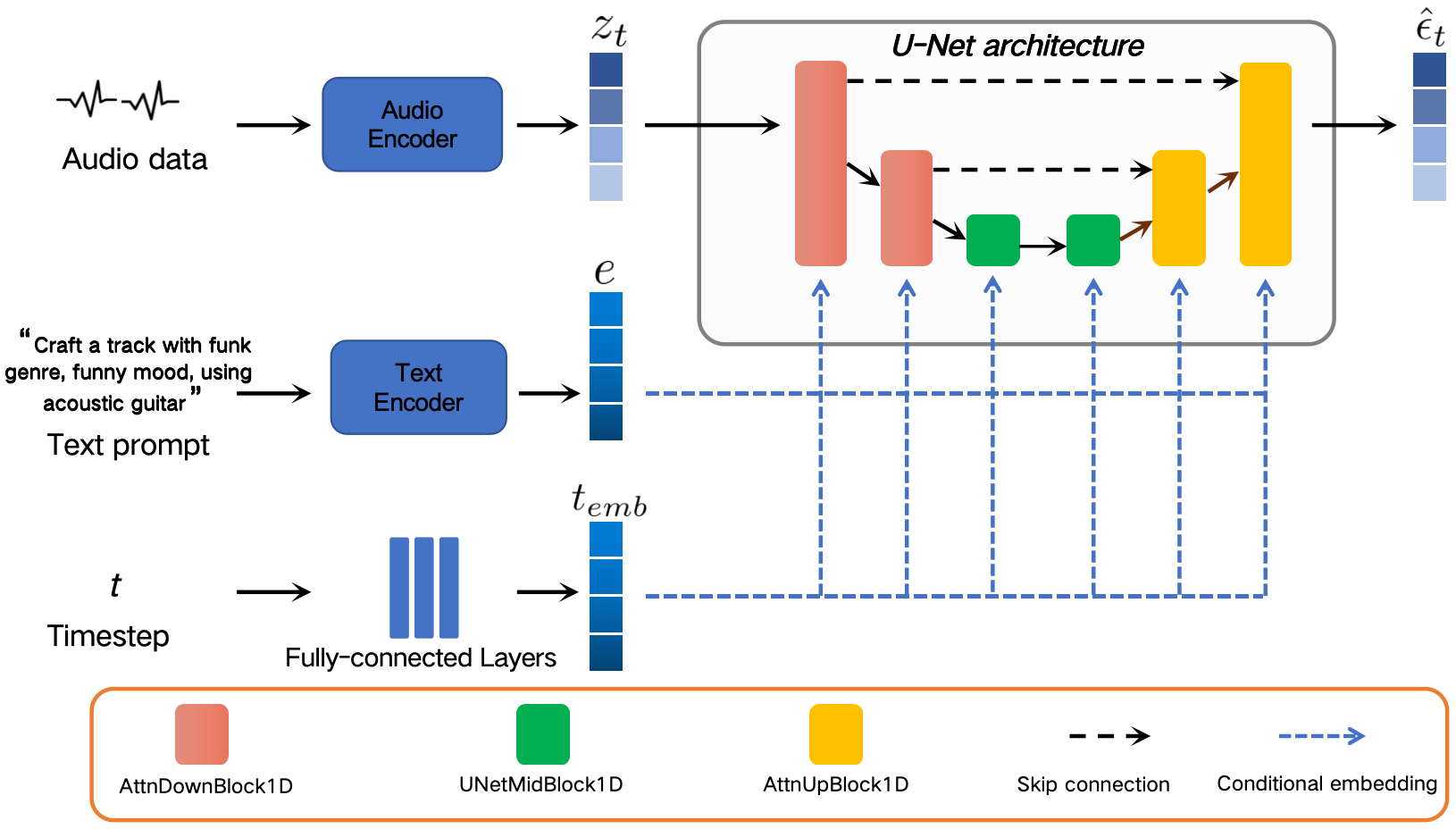}
    \caption{
The U-Net architecture used in Jen-1.
    }
    \label{fig:unet}
\end{figure}

In Jen-1~\cite{li2023jen}, the diffusion model $\theta$ operates in the audio latent space, predicting the noise $\widehat{\epsilon}_t=\epsilon_\theta(z_t, e, t)$ and iteratively denoising from Gaussian noise to generate the final latent $\widehat{z}_0$. The variable $e$ represents the embedding of conditioning inputs, such as text prompts, guiding the generation process.
Similar to LDM, Jen-1 utilizes a U-Net architecture~\cite{ronneberger2015u} as the backbone for its diffusion process. Specifically adapted for audio data, Jen-1 replaces the 2D convolutions used in image processing with 1D convolutions tailored for audio latent representations. The model consists of a sequence of blocks, including AttnDownBlock1D, UNetMidBlock1D, and AttnUpBlock1D, which integrate residual 1D convolutional layers with cross-attention transformers~\cite{vaswani2017attention}. The overall architecture is depicted in Figure~\ref{fig:unet}.

\begin{figure*}[tbhp]
\centering
\includegraphics[width=0.73\textwidth]{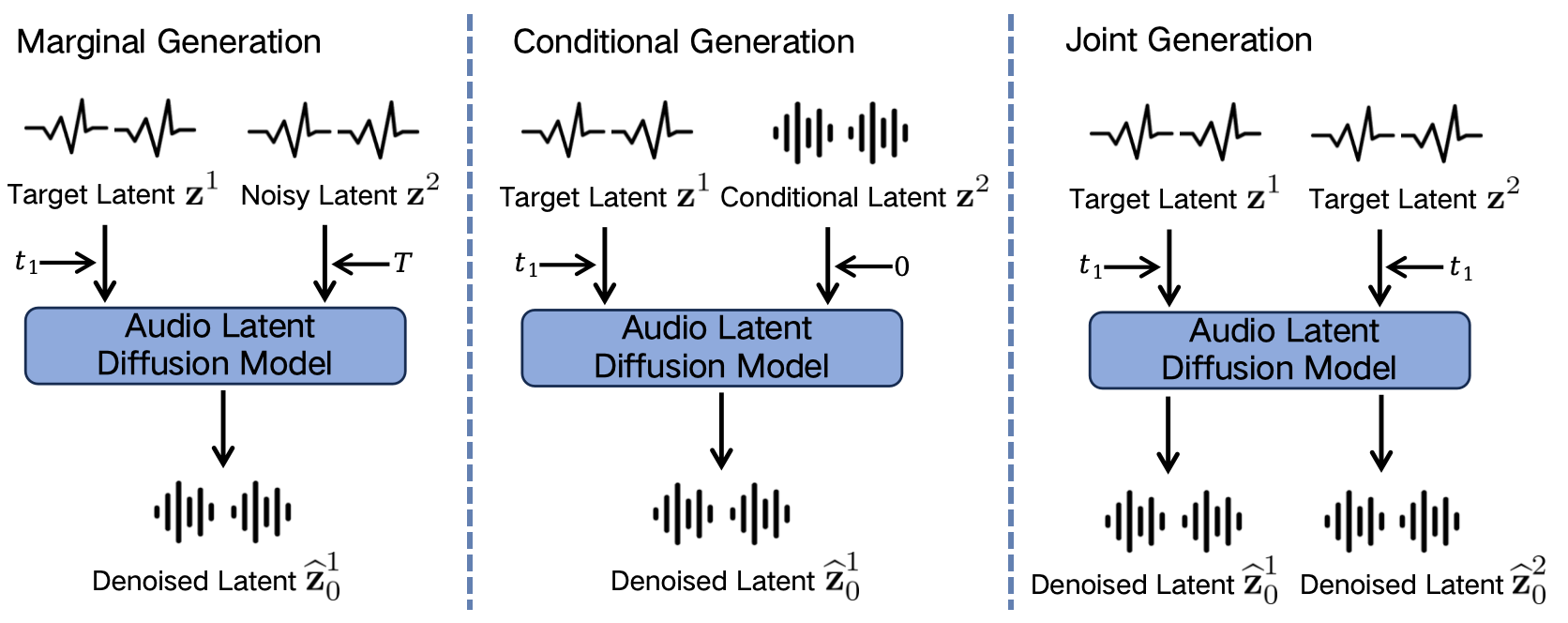}
\caption{
 Illustration of three generation modes using independent timesteps as indicators. In Marginal Generation, non-target track latents are fixed as Gaussian noise (timestep $T$) to minimize their impact on the target track's latent. Conditional Generation designates a timestep of $0$ for a conditional track, guiding the target track's generation. Joint Generation synchronizes multiple target tracks by sharing the same timestep $t$, allowing for coordinated denoising from $T$ to $0$.
}
\label{fig:t-illustration}
\end{figure*}

\section{Method}
\subsection{Multi-track Music Generation}
\label{sec:model}
To enable JEN-1 Composer to handle multi-track input and output for joint modeling, we make several critical modifications to Jen-1's single-track architecture. As elaborated below, the input-output representation, timestep vectors, and prompt prefixes are adapted to fit multi-track distributions efficiently using a single model.

\noindent\textbf{Multi-track Input-Output Representation.}
We extend the single-track input paradigm of Jen-1 to accommodate multi-track inputs, denoted as $\mathbf{X}=\left[x^1_0, \ldots, x^K_0\right]$, where $x^i_0$ represents the $i$-th track and $K$ denotes the total number of tracks.
Each track undergoes encoding into the audio latent space, yielding latent representations $z^i_0 = f_\phi(x^i_0) \in \mathbb{R}^{D \times S^{\prime}}$ prior to being inputted to the audio latent diffusion model.
These latent representations are concatenated along the channel dimension to form the input latent variables $\mathbf{Z} \in \mathbb{R}^{KD \times S^{\prime}}$.
During inference, the output of the audio latent diffusion model is split into $K$ tracks, with each denoised latent variable reconstructed into waveform via the pre-trained audio decoder, denoted as $\widehat{x}_0^i =g_{\psi}(\widehat{z}^i_0)$.
The extension of the input-output representation to multi-track enables explicit modeling of inter-track dependencies and consistency, crucial for high-quality multi-track generation that is absent in single-track models.

\noindent\textbf{Individual Timestep Vectors.}
Introducing separate timesteps for each track not only provides precise control over the generation process but also enables unified distribution modeling. This is achieved by extending the scalar timestep $t$ in Jen-1 to a multi-dimensional vector $\mathbf{T}=\left[t_1,\ldots, t_K\right]$. Each $t_i$ determines the corresponding latent variable $z^i$ in $\mathbf{Z}=\left[z^1,\ldots, z^K\right]$ according to the diffusion forward process defined in Equation~(\ref{latent:diff}). In the diffusion model, these timesteps are independently learned for each track and concatenated to form the conditional embedding. The process is formalized as follows:
\begin{equation}
\label{defz}
    z^i = 
\begin{cases} 
z^i_T, & \text{if }\ t_i = T \\
z^i_0, & \text{if }\ t_i = 0  \\
z^i_t, & \text{if }\ 0 < t_i < T 
\end{cases}
\end{equation}

For $k\geq 2$ target generation tracks, we adopt a uniform timestep $t$ across these tracks, mirroring the modeling of their joint probability distribution. Specifically, if $k<K$, two modes are considered for the remaining $(K-k)$ tracks. Firstly, the conditional generation mode sets all timesteps to 0, representing latent variables corresponding to original waveforms, akin to conditional generation. Here, the corresponding latent variables and timestep vectors are denoted as $\mathbf{Z}_c$ and $\mathbf{T}_c$, respectively. Secondly, the unconditional generation mode involves fixing all non-target timesteps to $T$, indicating perturbation of corresponding latent variables to approximate Gaussian random noise, akin to marginal generation. Correspondingly, the latent variables and timestep vectors are labeled as $\mathbf{Z}_m$ and $\mathbf{T}_m$. 
During loss computation, emphasis is laid solely on the channels corresponding to the target tracks, while inference benefits from the Classifier-Free Guidance (CFG) technique~\cite{ho2022classifier}. Specifically, if $k<K$, the alignment across tracks and generation quality are enhanced via the expression:
\begin{equation}
\label{eq.cfg}
    \widehat{\epsilon} = \left(1-\lambda \right)\epsilon_\theta\left(\mathbf{Z}_m, e, \mathbf{T}_m\right) + \lambda \epsilon_\theta\left(\mathbf{Z}_c, e, \mathbf{T}_c\right),
\end{equation}
where $\lambda$ denotes the guidance scale. Illustrations presented in Figure~\ref{fig:t-illustration} showcase diverse scenarios of a straightforward two-track generation task, providing additional clarity on the concept of achieving unified distribution modeling through the manipulation of timestep vectors.

\subsection{Integrating Task Tokens as Prefix Prompts}
\label{sec:prompt}
We enhance traditional text prompts, which typically describe the musical content and style, by integrating task-specific tokens as prefix prompts. These tokens act as explicit directives, similar to command flags in programming, providing clear and concise instructions regarding the generation task at hand. By using specific prefixes like ``[bass \& drum generation]", we direct the model's focus to the production of target tracks, such as bass and drums. This approach not only specifies the generation objectives but also significantly diminishes ambiguity, thus improving both the fidelity and relevance of the generated content.

\subsection{Progressive Curriculum Training Strategy}
\label{sec:train}
We introduce a progressive curriculum training strategy designed to systematically enhance the model's capability to generate coherent multi-track audio sequences while accommodating varying levels of conditioning and noise injection. This strategy includes curriculum decay and task allocation, a strategic sampler for conditional and marginal generation modes, and self-bootstrapping training to improve model generalization.

The training begins with single-track text-to-music generation, establishing a strong foundation for the model. As the model advances, more complex multi-track tasks are gradually introduced. This progression is carefully managed by reducing the sampling probabilities of simpler tasks, allowing the model to develop the ability to generate harmonically aligned tracks across multiple channels. Each task involves multi-track audio input and output, with latent representations configured as described in Equation (\ref{defz}). 
During this phase, the model's learning is focused on critical aspects by computing losses only for target tracks, while non-target tracks are masked. This structured approach facilitates efficient learning, enabling the model to generate high-fidelity audio compositions.

\noindent\textbf{Curriculum Decay and Task Allocation.} 
The curriculum starts with single-track ($k=1$) tasks, focusing on conditional generation using other tracks as signals or simpler marginal generation tasks. As training progresses, the focus shifts towards multi-track generation ($2 \leq k < K$), with increased sampling probabilities for more complex tasks over time. Ultimately, the curriculum incorporates joint generation tasks ($k=K$) driven solely by text prompts.

\noindent\textbf{Sampler for Conditional and Marginal Generation.}
A strategic sampler is employed when fewer target tracks are generated than available ($k < K$). The sampler assigns non-target tracks a timestep of either $0$ or $T$: with probability $p_1$, a timestep of $0$ is chosen to encourage conditioning, and with probability $1-p_1$, a timestep of $T$ is selected for non-conditioned generation. This approach allows the model to effectively learn both conditional and marginal generation, preparing it for CFG technique implementation during inference and enhancing its overall performance in generating coherent music tracks.

\noindent\textbf{Incorporation of Self-Bootstrapping Training.}
In later training stages, self-bootstrapping is introduced with a probability $p_2$ to improve generalization and align with the Human-AI co-composition workflow. During this phase, tracks generated by a teacher model---using an exponential moving average of the model's parameters---replace a portion of the ground truth-aligned conditional input tracks. This technique refines the model's alignment and synchronization capabilities, expands the training dataset, and enhances generalization, which is crucial for performance in real-world, interactive environments.

\begin{algorithm}[tb]
\caption{Human-AI Co-composition Workflow}
\label{algo:1}
\begin{algorithmic}[1]
\STATE \textbf{Input:} Text prompt, user-provided tracks $\mathbf{S}$ (optional)
\STATE \textbf{Output:} Set of selected and refined tracks $\mathbf{S}$

\STATE $e \leftarrow$ Embedding of the given prompt
\WHILE{$\mathbf{S}$ is empty}
    \STATE \textit{\# Joint Generation}
    \STATE $(\widehat{x}^1_0, \ldots, \widehat{x}^K_0) \leftarrow$ Model.GenerateTracks$(e)$
    \STATE $\mathbf{S} \leftarrow$ User.selectAndRefineTracks$(\widehat{x}^1_0, \ldots, \widehat{x}^K_0)$
\ENDWHILE

\WHILE{not all $K$ tracks are satisfactory}
    \STATE \textit{\# Using the CFG technique defined in Equation~(\ref{eq.cfg})}
    \STATE $(\widehat{x}^1_0, \ldots, \widehat{x}^K_0) \leftarrow$ Model.GenerateTracks$(\mathbf{S}, e)$
    \STATE \textit{\# Update $\mathbf{S}$}
    \STATE $\mathbf{S} \leftarrow \mathbf{S}\ \cup$ User.selectAndRefineTracks$(\widehat{x}^1_0, \ldots, \widehat{x}^K_0)$
\ENDWHILE
\end{algorithmic}
\end{algorithm}

\subsection{Human-AI Co-composition Workflow}
\label{sec:infer}
During inference, our model supports the conditional generation of multiple tracks given $0$ to $K-1$ input tracks. To facilitate Human-AI collaborative music creation, we propose an interactive generation procedure, outlined in Algorithm~\ref{algo:1}.
The proposed interactive inference approach effectively integrates human creativity with AI capabilities, enabling a collaborative music generation process. This workflow offers three primary benefits:
\begin{itemize}
\item \textbf{Enhanced Refinement.} The iterative feedback mechanism allows users to progressively refine each track, enabling nuanced improvements that purely AI-driven generation may struggle to achieve. By selecting and refining satisfactory tracks, users help steer the generation process toward desired outcomes, filtering out low-quality outputs.
\item \textbf{Alignment with Human Aesthetics.} The interaction between human creators and the model enhances the AI's understanding of human aesthetic preferences and sound quality standards. This ongoing collaboration ensures that the generated tracks align more closely with artistic intent and professional expectations.
\item \textbf{Creative Control and Engagement.} The collaborative experience empowers human producers, providing a sense of control over the creative process. By balancing AI-driven generation with human input, the workflow ensures that both improvisation and structural coherence are maintained, enabling the realization of creative visions with AI assistance.
\end{itemize}

\section{Experiment}
\subsection{Experimental Setting}
We conducted extensive experiments to evaluate the capabilities of JEN-1 Composer, focusing on its performance across various dimensions to understand its potential in real-world applications.

\noindent\textbf{Dataset Setup.}
Like Diff-A-Riff~\cite{nistal2024diff}, we prioritize high-quality audio generation suitable for professional use, which requires training on proprietary datasets due to their superior sound quality compared to open-source MIDI-based data. For JEN-1 Composer, we utilized an 800-hour private studio recording dataset comprising five temporally aligned tracks: bass, drums, instrument, melody, and the final mix. Each track is annotated with metadata, including genres (\textit{e.g.}, blues, folk), instruments (\textit{e.g.}, guitar, piano), moods (\textit{e.g.}, cheerful, romantic), tempo, keywords, and themes. The dataset is divided into 640 hours for training and 160 hours for testing. Tracks are randomly sliced into aligned segments of varying lengths to ensure the model learns inter-track dependencies, enabling it to generate cohesive, high-quality multi-track music guided by text prompts.

\begin{table*}[tb!]
\centering
\scriptsize
\begin{tabular}{lccccc|c}
\toprule
   \multicolumn{1}{c}{}& \multicolumn{5}{c}{\footnotesize\textsc{Clap}$ \uparrow$} & \multicolumn{1}{c}{\footnotesize\textsc{RPR}$ \uparrow$}\\ \cmidrule{2-7}
\textsc{Methods}       & \textsc{Bass}     & \textsc{Drums}  & \textsc{Instrument} & \textsc{Melody} & \textsc{Mixed} & \textsc{Mixed}\\
\midrule
MusicLM & $0.16$ & $0.17$ & $0.23$ & $0.28$ & $0.28$ & $27\%$\\
MusicGen & $0.17$ & $0.15$ & $0.25$  & $0.33$   & $0.35$ & $36\%$ \\
Jen-1 & $0.19$ & $0.16$ & $0.29$  & $0.32$  & $0.36$ & $40\%$ \\
\midrule
\brow \textbf{JEN-1 Composer}    & $\mathbf{0.21}^{\star}$   & $\mathbf{0.18}^{\star}$  & $\mathbf{0.29}$  & $\mathbf{0.36}^{\star\star}$  & $\mathbf{0.39}^{\star\star}$  & $\mathbf{-}$ \\
\bottomrule
\end{tabular}
\caption{
Comparison of mixed-track text-to-music generation. 
 $^{\star\star}$ and $^{\star}$ represent significance level $p$-value $< 0.01$ and $p$-value $<0.05$ of comparing JEN-1 Composer with Jen-1.
Wherever possible, we use open-source models, and for MusicLM, we employ the publicly available API.
}  
\label{tab:sota-performance}
\end{table*}

\noindent\textbf{Evaluation Metrics.}
Our evaluation employs both quantitative and qualitative metrics to assess the model's performance.

\textbf{Quantitative Evaluation.}
We use the CLAP score~\cite{elizalde2023clap} to evaluate how well the generated music aligns with the intended semantic content of the text prompts. Higher CLAP scores indicate better alignment. In our study, we compute these scores for both individual tracks and the aggregated mixed track (MIXED CLAP) to assess the effectiveness of JEN-1 Composer in adhering to textual descriptions. For comparison models, we apply Demucs~\cite{defossez2021hybrid,rouard2022hybrid} to separate tracks before calculating their individual CLAP scores. 
We also use Fr\'{e}chet Audio Distance (FAD)~\cite{roblek2019fr} as metric. We evaluated the quality of our generated mixed audio by comparing it to mixed audio from both our proprietary dataset and the public Slakh2100 dataset~\cite{manilow2019cutting} using the FAD metric (lower is better), computed with VGGish embeddings~\cite{hershey2017cnn}. This comparison was done without fine-tuning, providing a zero-shot evaluation. 
This approach ensures a fair and consistent evaluation of the contextual relevance and musical fidelity of the generated outputs across different methods.

\textbf{Qualitative Evaluation.}
We employ the \textbf{R}elative \textbf{P}reference \textbf{R}atio (RPR)
to capture human judgment of audio quality. Multiple raters evaluate pairs of audio samples---one generated by JEN-1 Composer and the other by a comparison model---without knowing the origin of each sample. Raters assess based on audio quality, coherence, harmony, and adherence to the text prompt. The RPR is recorded as a percentage, where $0\%$ indicates no preference for JEN-1 Composer's output, and $100\%$ indicates complete preference. This metric captures subjective preferences, providing insight into the perceived quality and effectiveness of the generated music.

By integrating CLAP scores, FAD, and RPR, we offer a comprehensive evaluation framework that balances objective alignment with subjective human perception, ensuring a thorough assessment of the model's strengths and areas for improvement.

\begin{table*}[tbp!]
\centering
\scriptsize
\begin{tabular}{lccccc|c}
\toprule
   \multicolumn{1}{c}{}& \multicolumn{5}{c}{\footnotesize\textsc{Clap}$ \uparrow$} & \multicolumn{1}{c}{\footnotesize\textsc{RPR}$ \uparrow$}\\ \cmidrule{2-7}
\textsc{Methods}       & \textsc{Bass}     & \textsc{Drums}  & \textsc{Instrument} & \textsc{Melody} & \textsc{Mixed} & \textsc{Mixed}\\
\midrule
baseline & $0.20$ & $0.18$  & $0.20$ & $0.28$ & $0.28$  & $16\%$ \\
\midrule
+ individual timestep vector  & $0.19$ & $0.18$  & $0.22$ & $0.32$ & $0.33$  & $20\%$ \\
+ curriculum training strategy  & $0.21$ & $0.17$  & $0.26$ & $0.35$ & $0.37$  & $35\%$ \\
\brow + interactive inference   & $\mathbf{0.21}$   & $\mathbf{0.18}$  & $\mathbf{0.29}$  & $\mathbf{0.36}$  & $\mathbf{0.39}$  & $\mathbf{-}$ \\
\bottomrule
\end{tabular}
\caption{Ablation studies. Starting from the baseline configuration, we incrementally modify the model to investigate the impact of each component. The baseline model involves only minimal modifications to the input and output channels of Jen-1 and employs a unified timestep for joint generation training.}
\label{tab:ablation}
\end{table*}

\noindent\textbf{Implementation Details.}
Our task involves generating four distinct audio tracks: bass, drums, instrument, and melody, alongside a synthesized mixed track. All tracks are high-fidelity stereo audio sampled at 48 kHz. We utilize the 48k version of the pre-trained EnCodec~\cite{defossez2022high}, resulting in a latent space representation of 150 frames per second, each with 128 dimensions. The volumes of individual tracks are adjusted to ensure consistent relative loudness before encoding.
For text encoding, we employ the pre-trained Flan-T5-Large model~\cite{chung2022scaling}, which provides robust capabilities for understanding and processing complex textual inputs. 
The architecture of JEN-1 Composer is built upon a 1D UNet backbone~\cite{ronneberger2015u}, with critical modifications to the Jen-1 model~\cite{li2023jen}, as outlined in Section~\ref{sec:model}. These key adjustments include the channel-wise concatenation of tracks and the expansion of the single-track timestep into a vector of four elements, enabling the model to effectively handle and generate cohesive multi-track music.

Training follows a progressive curriculum strategy, as detailed in Section~\ref{sec:train}. Initially, the probability for single-track generation tasks is set to $1/K$, with $K=4$, where each track is independently considered as a target track. As training progresses, these probabilities gradually decay, allowing for the introduction of more complex multi-track generation tasks. Eventually, all task types are covered, with the probability for each generation scenario (whether a track is a target track or not) set to $1/(2^K-1)$. Sampler settings for conditional and marginal generation are optimized with $p_1=0.8$. After 300 epochs, self-bootstrapping training is introduced with a probability $p_2=0.5$. We determined the optimal value for the guidance scale parameter $\lambda=7$ through a grid search. Training was conducted on two NVIDIA A100 GPUs, with hyperparameters including the AdamW optimizer~\cite{loshchilov2017decoupled}, a linear decay learning rate starting at $3\textit{e-}5$, a batch size of 12 per GPU, and optimization settings of $\beta_1=0.9$, $\beta_2=0.95$, weight decay of 0.1, and a gradient clipping threshold of 0.7.

\subsection{Comparison with State-of-the-art Methods}
We conduct a comprehensive comparison with leading text-to-music generation models, including MusicLM~\cite{agostinelli2023musiclm}, MusicGen~\cite{copet2023simple}, and Jen-1~\cite{li2023jen}, which primarily focus on single-track generation. In contrast, JEN-1 Composer supports multi-track generation with flexible conditional control, enabling track-wise generation and enhanced inter-track alignment.
As shown in Table~\ref{tab:sota-performance}, JEN-1 Composer achieves superior CLAP scores across individual tracks and the mixed track, highlighting its ability to generate music that closely adheres to text prompts while maintaining inter-track coherence. Additionally, the RPR metric results confirm a strong user preference for JEN-1 Composer’s outputs, demonstrating its effectiveness in delivering high-quality compositions.

To evaluate multi-track generation quality, we performed a zero-shot comparison on the Slakh2100 dataset~\cite{manilow2019cutting}, without any fine-tuning. As shown in Table~\ref{tab:sota-performance2}, JEN-1 Composer achieves competitive FAD scores compared to StemGen~\cite{parker2024stemgen} and MSDM~\cite{marianimulti}. Notably, our method, which infers directly on the Slakh2100 test set, achieves a FAD of 4.04, outperforming StemGen (4.30) and significantly surpassing MSDM (6.55). This indicates the strong generalization capability of our model, which can be attributed to the high-quality proprietary dataset used for training. Moreover, JEN-1 Composer achieves an even lower FAD of 3.76 on our proprietary dataset. This result is reasonable given that our dataset comprises high-fidelity studio recordings, as opposed to MIDI-rendered audio typically used in other studies. These findings emphasize the importance of high-quality training data and further demonstrate the robustness and audio fidelity of JEN-1 Composer.

\begin{table}[tbhp]
\centering
\scriptsize
\begin{tabular}{lcc}
\toprule
\textsc{Methods}           & \textsc{Datasets}      & FAD$\downarrow$ \\
\midrule
StemGen                  & Slakh                 & 4.30         \\
MSDM                     & Slakh                 & 6.55         \\
\midrule
\brow JEN-1 Composer           & Slakh                 & 4.04         \\
\brow JEN-1 Composer           & Proprietary           & 3.76         \\
\bottomrule
\end{tabular}
\caption{Comparison of multi-track generation quality. Results for StemGen and MSDM are taken directly from their original papers.}
\label{tab:sota-performance2}
\end{table}

\subsection{Ablation Studies}
Our ablation studies underscore the significance of each component within the JEN-1 Composer framework. As summarized in Table~\ref{tab:ablation}, we started with a baseline model featuring a four-track input-output configuration inspired by Jen-1~\cite{li2023jen}, and incrementally introduced our proposed enhancements. Using individual timestep vectors for each track was pivotal in effectively modeling both marginal and conditional distributions, resulting in notably higher CLAP scores for individual tracks.

The progressive curriculum training strategy, which transitions from simpler conditional generation to complex joint modeling tasks, further improved performance, particularly for intricate tracks like melody and instrument. Moreover, integrating an interactive Human-AI co-composition workflow yielded the highest mixing quality. This design allowed the model to alternate flexibly between generation modes and incorporate multiple human inputs as additional conditional signals. For example, the model could initially generate melody and instrument tracks, then leverage these as guidance to condition the subsequent generation of drums and bass, ensuring greater coherence across tracks.

In summary, the careful orchestration of these components enables JEN-1 Composer to achieve exceptional quality and flexibility in multi-track music synthesis.

\section{Conclusion}
In this study, we introduce JEN-1 Composer, a text-to-music framework for multi-track audio generation that enhances user interaction through intuitive, direct audio input. By moving beyond the limitations of symbolic and spectrum-based methods, JEN-1 Composer simplifies the creative process while preserving rich audio fidelity. The model not only leverages text prompts as conditional control signals but also integrates multi-track coherence as an additional guiding signal to ensure harmonious and well-aligned generation. Through curriculum training, it progressively learns to model marginal, joint, and conditional distributions, enabling the generation of both simple and complex multi-track compositions.

Despite these advancements, achieving professional music standards remains a challenge. Future improvements will require deeper integration of engineering, design, and the arts, alongside enhanced data collection and annotation processes, to further refine the model’s control, coherence, and output quality.

\section*{Ethical Statement}
In developing JEN-1 Composer, we strictly adhered to ethical standards, using only authorized data and complying fully with copyright and licensing agreements. JEN-1 Composer is designed to enhance, not replace, human creativity, promoting Human-AI collaboration to push artistic boundaries. We are committed to the ethical use of AI, opposing any misuse for unauthorized reproduction or plagiarism. Our approach prioritizes creators' rights and encourages responsible AI integration in the creative process. This manuscript was polished and proofread with the assistance of ChatGPT.

\bibliography{aaai25}

\end{document}